\documentclass[aps,pra,onecolumn,amsfonts,amssymb]{revtex4}

\usepackage{graphics}
\usepackage{epsfig}
\usepackage{amsmath}
\usepackage{amsthm}

\usepackage{amssymb, amscd}
\usepackage{graphicx}
\usepackage{subfigure}
\usepackage{enumerate}

\begin{document}

\title{Transmitting qudits through larger quantum channels}

\author{Grigori G. Amosov}
\email{gramos@mail.ru}
\affiliation{PN Lebedev Physical Institute, Leninskii Pr. 53, Moscow
119991, Russia $ $}

\author{Stefano Mancini}
\email{stefano.mancini@unicam.it}
\affiliation{Dipartimento di Fisica, Universit\`a di Camerino, 
I-62032 Camerino, Italy\\
\& INFN Sezione di Perugia, I-06123 Perugia, Italy}

\author{Vladimir I. Manko}
\email{manko@sci.lebedev.ru}
\affiliation{PN Lebedev Physical Institute, Leninskii Pr. 53, Moscow
119991, Russia}

\date{\today}

\begin{abstract}
We address the problem of transmitting
states belonging to finite dimensional Hilbert space 
through a quantum channel associated with
a larger (even infinite dimensional) Hilbert space.
\end{abstract}

\maketitle

\section{Introduction}

By a quantum channel is intended \cite {Hol, Keyl} a completely positive and
trace-precerving linear map $\Phi :\sigma (H)\to
\sigma (H)$, where $\sigma (H)$ is a set of states
in a Hilbert space $H$ with $dim H=n\leq +\infty$. Even for a finite
dimension $n<+\infty $ there is a number of difficult
problems concerning the construction of optimal transmission of
the information through a quantum channel. The infinite dimensional
case we are especially interested in, has
additional particular features \cite {Shir}. In the present paper we discuss
how a set of states on the finite dimensional Hilbert space can
be transmitted through a quantum channel associated with
the infinite dimensional Hilbert space. In particular, we investigate whether 
encoding a qudit in a larger space could be useful to better protect it from the channel (noise) action without resorting to any particular decoding (recovery) scheme at the output.

The paper is organized as follow. In Sec. II
the notion of subchannel is introduced. In Sec. III
we study the phase and the amplitude damping channels within this context.
Sec. IV is for conclusions.

\section{Invariant hulls and subchannels}

Given a subspace
$K\subset H$, $dim K=d<n$, we denote by $\sigma  (K)$
the convex envelope of pure states $|\xi\rangle\langle\xi|\subset \sigma (H),\
\xi\in K$. One can define a linear map
$\Psi :\sigma (K)\to \sigma (K)$ by the formula
\begin{equation}\label{Tr1}
{\rm Tr}(x_2\Psi (x_1))={\rm Tr}(x_2\Phi (x_1)),\quad x_i\in \sigma (K), \quad i=1,2.
\end{equation}
Substituting $x_1=|\xi\rangle\langle\xi |$ into Eq.(\ref{Tr1})
we obtain $\langle\xi|\Psi (x_2)|\xi\rangle=\langle\xi|\Phi (x_2)|\xi\rangle$. It follows
that $\Psi (x)=P_K\Phi (x)P_K$, where $P_K$ is a projection on
the subspace $K$ and $x\in \sigma (K)$. Hence $\Psi $ is a
completely positive map.
 If $\Phi $ is unital,
i.e. it preservs the chaotic state $\Phi (\frac {1}{n}I)=\frac
{1}{n}I$ (with $I$ the identity in $H$), then $\Psi $ satisfies the property
$\Psi (\frac {1}{d}P_K)=\frac {1}{d}P_K$.

Consider the set of states
\begin{equation}\label{Im}
Im_K\Phi \equiv\{x\in \sigma (H)\ |\ \exists y\in \sigma (K)\ :\ \Phi (y)=x\}.
\end{equation}
If $Im_K\Phi \subset \sigma (K)$, then we shall call
$\sigma (K)$ {\it an invariant hull}
of the channel $\Phi $.
In that case we get $\Psi =\Phi |_{\sigma (K)}$, where $\Phi |_{\sigma (K)}$ stands the restriction of  $\Phi$ to inputs in ${\sigma (K)}$.
Because it implies that $\Psi $ is trace-preserving, we shall call
$\Psi $ {\it a subchannel} of $\Phi $.

We denote by $B(H)$ the algebra of all bounded operators in $H$.
Due to the Kraus decomposition\footnote{This terminology arose because of the Kraus' book \cite{Kraus} where the decomposition appeared, however it was first proposed in Ref.\cite{Sud}.} for the channel $\Phi $ there exist
a set of operators $E=\{E_i\in B(H),\ 1\leq i\leq k\leq n^2 \}$
such that
\begin{equation}\label{Kd1}
\Phi (x)=\sum \limits _{i=1}^kE_ixE_i^*,\quad x\in \sigma (H).
\end{equation}
where $E_i^*$ stands for the adjoint of $E_i$.

\bigskip

{\bf Example 1.} Consider the phase damping channel defined with $n=+\infty$ through the decomposition
\begin{equation}\label{pd}
\Phi (x)=\sum \limits _{i=0}^{+\infty }E_ixE_i^*,
\end{equation}
where
\begin {equation}\label{Epd}
E_{i}=\sum_{k=0}^{\infty}
\frac{\left[k\sqrt{-2 \ln \eta}\right]^{i}}
{\sqrt{i!}}\left[\eta\right]^{k^{2}}\;
|k\rangle\langle k|\,.
\end{equation}
Here $|k\rangle$, $k=0,1,2,\ldots$, are Fock states and
the parameter $\eta$ describes the damping
(it can be written as $\eta=e^{-\gamma t}$ with
$\gamma$ the damping rate and $t$ the transmission time).
The property $\Phi (|k\rangle\langle k|)=|k\rangle\langle k|$ guarantees that
any subspace $K$ generated by the vectors
$|k_0\rangle,\dots ,|k_{d-1}\rangle$ determines the invariant hull
of the phase damping channel.

\bigskip

{\bf Example 2.} Consider the amplitude damping channel defined with $n=+\infty$ through the decomposition
\begin{equation}\label{ad}
\Phi (x)=\sum \limits _{i=0}^{+\infty }E_ixE_i^*,
\end{equation}
where
\begin{equation}\label{Ead}
E_i=\sum \limits _{k=i}^{+\infty }\sqrt {C_k^i}\,[\eta]^{(k-i)/2}[1-\eta ]^{i/2}|k-i\rangle\langle k|,
\end{equation}
with
\begin{equation}\label{C}
C_k^i=\frac{k!}{(k-i)!i!}
\end{equation}
Here again  $|k\rangle$, $k=0,1,2,\ldots$, are Fock states and
the parameter $\eta$ describes the damping.
For a given dimension $d$ there exists the invariant hull for $\Phi$.
In fact, if $K$ is generated by a collection of vectors $|0\rangle,\ldots,|d-1\rangle$, 
then $\sigma (K)$ is an invariant hull for $\Phi $.

\bigskip

{\bf Example 3.} Consider the depolarazing channel defined as
$\Phi (x)=px+(1-p)\frac {1}{n}I$, with $0\le p\le 1$.
Notice that if $K\neq H$, then the chaotic state $\frac {1}{n}I\notin
\sigma (K)$. Hence the depolarizing channel has no invariant hull
if $d<n$ because
$Im_K\Phi \notin \sigma (K)$ under this condition.

\bigskip
\bigskip

For any channel $\Phi$ in the Hilbert space $H$ one can define the conjugate unital completely
positive map $\Phi ^*$ as follows
\begin{equation}\label{Tr2}
{\rm Tr}(x_1\Phi ^*(x_2))={\rm Tr}(\Phi (x_1)x_2),\quad x_1\in \sigma (H),\ x_2\in B(H).
\end{equation}
Because $\Phi $ is trace-preserving, we obtain that $\Phi ^*$ is
unital. Moreover to check that the map $\Phi $ is trace-preserving it
is sufficient to look whether $\Phi ^*$ is unital or not.
Equation (\ref{Kd1}) allows us to extend $\Phi $ from $\sigma (H)$
to $B(H)$. For the conjugate map $\Phi ^*$ we get
\begin{equation}\label{Kd2}
\Phi ^*(x)=\sum \limits _{i=1}^kE_i^*xE_i,\quad x\in B(H).
\end{equation}

Now let $\Psi (x)=P_K\Phi (x)P_K,\ x\in \sigma (K)$. The map $\Psi $ is
a subchannel of $\Phi $ iff it is trace-preserving. To fulfill this
property it needs that $\Psi ^*$ is unital in the sense
$\Psi ^*(P_K)=P_K$. On the other hand, it takes place iff
\begin{equation}\label{PK}
P_K\Phi ^*(P_K)P_K=P_K.
\end{equation}
Hence, the problem of
searching for subchannels of $\Phi $ is equivalent to
the problem of describing the algebra of fixed elements for
the map $P_K\Phi ^*(\cdot )P_K$.

If $\Psi $ is a subchannel of $\Phi $ and $dimK=2$, then we shall
say that $\Psi $ is a \textit{qubit subchannel} of $\Phi $. 
Let 
\begin{equation}\label{psi12}
|\psi\rangle
=\cos(\frac {\theta }{2})|\psi _0\rangle+e^{i\phi }\sin (\frac {\theta
}{2})|\psi _1\rangle,\quad\rho =|\psi \rangle\langle\psi |,
\end{equation} 
with $|\psi_0\rangle$, $|\psi_1\rangle$ spanning $K$.
A way to see how faithfully the state $\rho$ is transmitted through the channel $\Phi$ is to 
consider the fidelity distance \cite{Keyl}
\begin{equation}\label{f}
f(\theta,\phi)={\rm Tr}(\rho, \Phi (\rho))
\equiv {\rm Tr}\left \{\sqrt {\sqrt {\rho }\,\Phi(\rho)\, \sqrt
{\rho } }\right \}=\langle\psi |\Phi(\rho) |\psi \rangle,
\end{equation}
and then average overall the Bloch sphere, to get  
\begin{equation}\label{calF}
{\mathcal F}= \frac {1}{4\pi}\int \limits _{0}^{2\pi}d\phi \int \limits
_{0}^{\pi}d\theta \sin(\theta)f(\theta ,\phi).
\end{equation}
It turns out that the fidelity of the channel $\Phi $ is equal to the fidelity of its
qubit subchannel $\Psi $ (with the subspace $K$ generated by the
vectors $|\psi _0\rangle$ and $|\psi _1\rangle$).  Following this way, we
can conclude that to estimate how well a channel $\Phi $
preserves a qubit state, one should consider all qubit
subchannels of $\Phi $.
 
 \section{Applications}

 \subsection{The phase damping channel}
 
  Suppose that $\Phi $ represents the phase damping channel defined in Example 1 of Sec. II. The
projectors $|k\rangle\langle k|$, $k=0,1,2,\ldots$, belong to the algebra of
fixed elements of $\Phi $. Hence, any subspace $K$ being a linear
envelope of the vectors $|k\rangle$ and $|s\rangle$, $k\neq s,$ generates the
unital qubit subchannel of $\Phi $. Notice that as consequence of Eqs.(\ref{pd}) and (\ref{Epd}) we have
\begin{equation}\label{Pkspd}
\Phi (|k\rangle\langle s|)=\eta ^{(k-s)^{2}}|k\rangle\langle s|.
\end{equation}
Thus, by referring to Eq. (\ref{psi12}) with $|k\rangle\equiv|\psi_0 \rangle$ and $|s\rangle\equiv|\psi_1\rangle$, we get
\begin{equation}\label{Tr3}
{\rm Tr}(\rho,\Phi (\rho))=\cos^{4}\left(\frac {\theta}{2}\right)+
\sin^{4}\left(\frac {\theta}{2}\right)+2\eta ^{(k-s)^{2}}\cos^{2}\left(\frac
{\theta}{2}\right)\sin^{2}\left(\frac {\theta}{2}\right),
\end{equation}
hence the fidelity 
 \begin{equation}\label{Fpd}
 {\mathcal F}=\frac {2}{3}+\frac {\eta
^{(k-s)^{2}}}{3}.
 \end{equation}
The maximum is achieved for $k$, $s$ contiguous natural numbers.

\subsection{The amplitude damping channel}

Suppose that $\Phi$ represents the amplitude damping channel defined in Example 2 of
Sec. II. Take two integer numbers $0\leq k\leq s$,
then as consequence of Eqs.(\ref{ad}), (\ref{Ead}) and (\ref{C}) we have
\begin{equation}\label{Pksad}
\Phi (|k\rangle\langle s|)=\sum \limits _{i=0}^k\sqrt {C_k^iC_s^i}\eta ^{\frac {k+s}{2}-i}(1-\eta )^i
 |k-i\rangle\langle s-i|.
\end{equation}
If $x=\sum \limits _{k=0}^{+\infty }x_{kl}|k\rangle\langle l|$, then for $y=\Phi (x)=
\sum \limits _{k=0}^{+\infty }\sum \limits _{l=0}^{+\infty }y_{kl}|k\rangle\langle l|$ we obtain
\begin{equation}\label{ykl}
y_{kl}=\sum \limits _{i=0}^{+\infty }\sqrt {C_{k+i}^iC_{l+i}^i}\eta ^{\frac {k+l}{2}}(1-\eta )^ix_{k+i\,l+i}.
\end{equation}
It follows from Eq.(\ref{ykl}) that $\Phi (x)= x $ iff $x=const |0\rangle\langle 0|$. 
 Hence, there is no
unital qubit subchannel of the amplitude damping channel. In fact, if the subspace $K$ generates an
invariant qubit of $\Phi $, then the subchannel $\Psi =\Phi |_{\sigma (K)}$ is unital only
if $\Phi (P_K)=P_K$, where $P_K$ is two-dimensional projection on $K$.
 
 \bigskip
 
 {\bf Example 5.}
Given a complex number $\alpha \in \mathbb{C}$ one can define the coherent state by the formula
\begin{equation}\label{coh}
|\alpha \rangle=e^{-\frac {|\alpha |^2}{2}}\sum \limits _{k=0}^{+\infty }\frac {\alpha ^k}{\sqrt {k!}}
|k\rangle.
\end{equation}
It follows from Eq.(\ref{Pksad}) and (\ref{coh}) that for any $\alpha ,\beta \in \mathbb{C}$ we get
\begin{equation}\label{Pab}
\Phi (|\alpha \rangle\langle\beta |)=|\sqrt {\eta }\alpha \rangle\langle\sqrt {\eta }\beta |\exp
 \left[(1-\eta )\left(-\frac {|\alpha |^2+|\beta |^2}{2}+\alpha \beta ^*\right)\right].
\end{equation}
 In particular,
\begin{equation}\label{Paa}
\Phi (|\alpha \rangle\langle\alpha |)=|\sqrt \eta \alpha\rangle\langle\sqrt \eta \alpha |
\end{equation}
for any coherent state $|\alpha \rangle\langle\alpha |$. Equation (\ref{Paa}) implies
that the Schr\"odinger cat states $|\psi _{\pm}\rangle= {\cal N}_{\pm}(|\alpha \rangle\pm |-\alpha \rangle)$ do not form an invariant qubit for
$\Phi $.
 
 \bigskip
\bigskip
 
Quite generally we can consider the qubit subchannel $\Psi $ of the amplitude
damping $\Phi $ generated by the vectors $|\psi_0\rangle, |\psi_1\rangle$ (basis for the qubit subspace $K$) given by  
\begin{equation}\label{enc}
|\psi_0\rangle=\sum \limits _{n=0}^{+\infty}c_{n}|n\rangle,\qquad |\psi_1\rangle=\sum \limits _{n=0}^{+\infty}d_{n}|n\rangle.
\end{equation}
The conditions
\begin{equation}\label{norm}
\sum \limits _{n=0}^{+\infty}|c_{n}|^{2}=1,\qquad \sum \limits
_{n=0}^{+\infty}|d_{n}|^{2}=1,
\end{equation}
and
 \begin{equation}\label{ort}
\sum \limits _{n=0}^{+\infty}{\overline c}_{n}d_{n}=0, 
\end{equation}
ensure the normalization of $|\psi_0\rangle$ and $|\psi_1\rangle$
and their orthogonality. Then, we consider a generic qubit state like in Eq.(\ref{psi12}).
Notice that the completely
positive map $\Psi=\Phi |_{\sigma(K)}$ is a non-unital qubit subchannel of $\Phi $ at least
if $c_{0}=1$ and $d_{1}=1$.  
 
From Eq.(\ref{calF}) we get the fidelity as
\begin{eqnarray}
{\mathcal F}&=&\frac {1}{6}\sum \limits _{k=0}^{+\infty}\sum \limits
_{n,m=k}^{+\infty}\sqrt
{C_{n}^{k}C_{m}^{k}}\,[\eta]^{(n+m-2k)/2}[1-\eta]^{k/2}\nonumber\\
&&\times
[c_{n}c_{m}(d_{m-k}{\overline d}_{n-k}+2c_{m-k}{\overline c}_{n-k})+d_{n}{\overline d}_{m}(c_{m-k}{\overline c}_{n-k}+2d_{m-k}{\overline d}_{n-k})\nonumber\\
&&\;\;\;\;+d_{n}{\overline d}_{n-k}{\overline c}_{m}c_{m-k}+c_{n}{\overline c}_{n-k}{\overline d}_{m}d_{m-k}].
\end{eqnarray}
Now we should maximize the fidelity overall possible choices of
$\{c_{n}\}$ and $\{d_{n}\}$. Clearely, there are no two
simultaneous and orthogonal eigenstates of $\Psi $, hence
the maximum of $\mathcal F$ can not be $1$.

If we use only the first two vectors $|0\rangle$ and $|1\rangle$ to
parametrize the qubit, then any choice of
$c_{0},c_{1},d_{0},d_{1}$ obeying the orthogonality condition
(\ref{ort}) gives us a correctly defined non-unital qubit subchannel
$\Psi=\Phi |_{\sigma(K)}$. In that case, the fidelity results 
\begin{equation}\label{Fad}
{\mathcal F}=\frac
{1}{2}+\frac {\eta}{6}+\frac{\sqrt{\eta}}{3}
\end{equation}
 and it does not depend on the choice of
$c_{0},c_{1},d_{0},d_{1}$.

One step ahead is to parametrize the qubit by the first three
vectors $|0\rangle,|1\rangle,|2\rangle$ as
\begin{eqnarray}
|\psi_0\rangle&=&\sin\alpha\cos\beta|0\rangle+\sin\alpha\sin\beta|1\rangle+\cos\alpha
|2\rangle,\\
|\psi_1\rangle&=&\sin\gamma\cos\delta|0\rangle+\sin\gamma\sin\delta|1\rangle+\cos\gamma|2\rangle,
 \end{eqnarray}
 with the condition
\begin{equation}
\sin\alpha\cos\beta\sin\gamma\cos\delta+\sin\alpha\sin\beta\sin\gamma\sin\delta
 +\cos\alpha\cos\gamma=0.
\end{equation}
Notice that we have skipped all relative phases because of the
symmetry of amplitude damping channel action. So, in practice we
only deal with three free parameters, $\alpha ,\beta ,\gamma
$. In that case, the fidelity is upper bounded by Eq.(\ref{Fad}) and such a bound is achieved with
 $\alpha =\gamma =\frac {\pi}{2}$ and
independently of $\beta $. The same takes place when parametrizing the qubit with more than three states.

Thus, it results clear that the fidelity is optimized by encoding the
qubit into the lowest two Fock states of the Hilbert space
(namely $|0\rangle$ and $|1\rangle$). This is in agreement with the physical arguments based on
energy (amplitude) damping. We conjecture that such a result can be 
extended from qubit to qudit subchannels. That is,
the optimal (in terms of fidelity) way to transmit a qudit through
a $n\geq d$ dimensional amplitude damping channel would be to encode it
into the lowest (in terms of energy) $d$ orthogonal states $|0\rangle,|1\rangle,\dots ,|d-1\rangle$.

\section{Conclusion}

We have formally addressed the problem of transmitting qudits through larger quantum channels by introducing the concepts of invariant hulls and subchannels. 
We have considered qudits encoded in the larger space of the channel 
without resorting to any particular decoding (recovery) scheme at the output.
After applying these arguments to specific examples, it comes out that to send a qubit through an infinite dimensional phase damping channel the best would be to encode it into two contiguous Fock states, while to send a qubit through an infinite dimensional amplitude damping channel the best would be to encode it into the two lowest Fock states. Although these results are derived from an information theoretic approach, they are in agreement with those coming from physical arguments.
These results can be generalized to qudits. 

Moreover, from the presented examples, it turns out that transmitting qudits in a channel of dimension greater than $d$ (even infinity) does not allows for a better fidelity, with respect to the case of a channel of dimension $d$.
Nevertheless, we believe that the extra space could be profitably exploited with suitable deconding, recovery procedures (see also Ref.\cite{presk}).  
From now we may argue that they cannot simply be completely positive trace preserving maps operating at the output of the channel as these are not able to decrease the distance between input and output states. Some possibilities will be addressed in future works.

\end{document}